\documentclass{article}
\usepackage{amssymb,latexsym}
\usepackage{amsmath}

\begin{document}
\title{Comment on Frank Wilczek's essay ``Total Relativity:  Mach 2004''}
\author{G. E. Hahne,
        P. O. Box 2748,
        Sunnyvale, California, 94087 USA}
\maketitle


\vskip 5pt
\begin{abstract}
     Frank Wilczek's essay ``Total Relativity:  Mach 2004''
(PHYSICS TODAY, April, 2004, page 10) cogently updates the status of the
intuitively compelling, but partly unrealized theory of, Mach's
principle.  I think, however, that an important consequence
of the principle has  been skipped:  that is, the possibility of
linear deformations of the internal structure of elementary particles from the
Minkowski structure imposed by the local gravitational (metric) field.  
[Note:  This Comment is similar
to one submitted to, but not published by, PHYSICS TODAY as a 
Letter to the Editor.]
\end{abstract}


      Einstein's field equations for gravitation 
(called General Relativity, although what is generalized
is not relativity but flat Minkowski spacetime to 
curved Riemann-Minkowski spacetime)
do realize Mach's principle in part:  The metric tensor field 
$g_{\mu\nu}(X)$, where 
\begin{equation}
(X)=(X^0,X^1,X^2,X^3)=(t,{\bf r})
\end{equation}
labels a point in space-time, is equivalent 
geometrically to
(i) a field of space-time volume measure $|\det(g(X))|^{1/2}dX^0dX^1dX^2dX^3$,
and (ii) a field of light cones in the tangent planes,
defined by the manifold of solutions to the $X$-dependent
quadratic equations (summation convention from 0 to 3 is on)
\begin{equation}
g_{\mu\nu}(X)\,V^\mu(X)\,V^\nu(X)\ =\ 0,
\end{equation}
where $V(X)$ is a vector in the tangent plane at $X$.
These fields are determined by the distribution of matter
everywhere in the universe, and by initial+boundary conditions on the fields. 

     Consider a point transformation in the spacetime manifold,
in the presence of a fixed, locally inertial coordinate system.
Then near the coordinate origin the transformation takes the form 
\begin{equation}
\bar{X}^\lambda= D^\lambda+E^{\lambda}_{\mu}X^\mu+
F^{\lambda}_{\mu\nu}X^\mu X^\nu 
+\text{higher-order terms.}\notag
\end{equation}
The four parameters in $D^\lambda$ represent uniform displacements 
in spacetime, the sixteen parameters in $E^{\lambda}_{\mu}$
include the six-parameter subgroup of homogeneous Lorentz
transformations, and the forty parameters in $F^{\lambda}_{\mu\nu}$
include the six-pa\-ram\-e\-ter nongroup family of nonrelativistic,
uniform linear accelerations and the leading terms
in a uniformly evolving rotation:
\begin{subequations}
\begin{align}
\bar{t}&=t,\notag \\
\bar{\bf r}&={\bf r}+(t^2/2){\bf a}+t({\bf \omega}\times{\bf r})
\ +\ \text{higher-order terms.}\notag
\end{align}
\end{subequations}

     We imagine a single elementary particle to be a
trajectory in the spacetime manifold, 
where the particle has, at each point in its trajectory,
an internal structure consisting
of a light cone, a space-time volume, and possibly other 
intrinsic attributes as spin angular momentum, electric charge, etc.
As the particle evolves along its dynamical trajectory,
its internal light cone and spacetime volume are,
in common contemporary theories, presumed implicitly to be constrained
to track precisely with the ambient structures established
by gravitation.  The principle of relativity
states, in effect, that the rest of the universe affords no local
opposition, or generalized restoring force, to the ``deformations''
of the particle afforded by the ten-parameter Poincar\'e group,
subject to the further constraint that, for a particle
with spin, the ``internal'' Lorentz boost is locked in
to the external one associated with its trajectory in spacetime;
a particle so deformed stays where it was put---Newton's First Law.
As originally proposed, Mach's principle asserts that
the deformation of a trajectory corresponding to a
linear acceleration is opposed by a restoring force
according to the formula ${\bf F}=-m{\bf a}$;  
uniformly evolving rotations give rise to internal
stresses and strains, as in Newton's bucket experiment (to be sure,
the second effect derives at least in part from the first).

     Although is is easy to effect the above-described special types of
second-order deformations in practice, it is plausible that a hypothetical
physical theory that realizes Mach's principle in the form
of ``total relativity'' should first
deal with the ten-parameter family of linear deformations
of the particle's internal structure
that is omitted when we confine our attention to the
Poincar\'e group.  Such deformations, according to
a total version of Mach's Principle, might occur if a particle's
internal structure were at least partly decoupled 
from the guiding field afforded by gravitation. 
These comprise (i) the one-parameter
subgroup of spacetime-isotropic dilations of a particle, and (ii) the
nine-parameter nongroup family of
anisotropic strain deformations of a particle.  Elements in (i),
which belong to the class of conformal transformations,
do not deform a particle's light cone, but deform its spacetime volume;
elements in (ii) deform its local light cone, but do not affect its
spacetime volume.  Both these types of deformation of a
particle are resisted so strongly by the rest of the universe
that they are apparently unphysical.  The restoring force exerted by
the universe to these deformations 
may not be infinite, however; it is
possible that, under presently undiscovered extreme conditions,
elementary particles respond with nonzero strain to
suitable forces. 
In other words, elementary particles might display additional
(discrete or continuous)
internal degrees of freedom, that are excited when intense isotropic
or anisotropic stresses are imposed on them.  A theory
of total relativity should then describe how it comes to be that
the guiding field in the form of gravitation is imperfect,
in that it is difficult, 
but not impossible, to make internal strain degrees of freedom
of an elementary particle to manifest themselves.

\end{document}